\documentclass{cimento}
\usepackage{amsmath,amssymb,graphicx} 

\title{Sill distribution: genesis and salient features}
\author{F.~Giacosa\from{ins:UJK}\thanks{francesco.giacosa@gmail.com}\ETC,
        \atque
V.~Shastry\from{ins:CEEM}\from{ins:IUB}\thanks{vanamalishastry@gmail.com}}
\instlist{\inst{ins:UJK} Institute of Physics, Jan Kochanowski University, ul.  Uniwersytecka  7,  P-25-406  Kielce,  Poland
  \inst{ins:CEEM} Center for  Exploration  of  Energy  and  Matter, Indiana  University, Bloomington,  IN  47403,  USA
  \inst{ins:IUB} Department of Physics, Indiana  University, Bloomington,  IN  47405,  USA}

\begin{document}
\maketitle

\begin{abstract}
   We present the so-called Sill distribution, both in the nonrelativistic and relativistic cases, as a natural and simple way to include the effect of threshold(s) on the energy line shapes of resonances. The Sill is correctly normalized (even for broad states), is continuous at threshold(s), does not require any modification to the `mass part', is easily extendable to the multichannel case, and can be applied to both mesons and baryons. Here, as a novel example, we employ the Sill to describe the resonance $\psi(3770)$.
\end{abstract}

The description of line shapes (or energy distributions) of unstable states is
an important element of both Quantum mechanics (QM) and Quantum Field Theory
(QFT) \cite{salam}. This is especially true for the study of short-lived hadrons, in which
broad states with relatively close thresholds are quite common \cite{pdg}. The aim of this work is to present the genesis and the main properties of a relatively simple distribution, called Sill distribution, originally put forward in Ref. \cite{Giacosa:2021mbz}.

The famous Breit-Wigner (BW) distribution \cite{ww,breit} (see also \cite{ceci} for modern applications),
\begin{equation}
d^{\text{BW}}(E)=\frac{\Gamma}{2\pi}\left[  (E-M)^{2}+\frac{\Gamma^{2}}%
{4}\right]  ^{-1}\text{ ,}%
\end{equation}
describes (in the non-relativistic limit) the line shape of a resonance
with mass $M$ and decay width $\Gamma$ that fulfills the normalization
$\int_{-\infty}^{+\infty}\mathrm{dE}d^{\text{BW}}(E)=1.$ According to BW, any
energy in the range $(-\infty,+\infty)$ is admissible, what is clearly a
nonphysical feature since any physical system has a minimal `threshold' energy
$E_{th}$, being the energy of the ground state ($E_{th}\geq0$ in QFT with
\ $E_{th}=0$ only if all the decay products of the resonance are massless
particles). One may include this feature by a simple rescaling
\begin{equation}
d^{\text{BW}}(E)\rightarrow Nd^{\text{BW}}(E)\theta(E-E_{th}),
\label{rescal}
\end{equation}
where $\theta(E)$ is the step function and $N\geq1$ is necessary to cure the
loss of normalization. While this procedure may be a good strategy in certain
applications, it is clear that it does not follow from a rigorous treatment of
the problem.

In general, the proper inclusion of the decay in a quantum context implies the
calculation of the self-energy $\Pi(E)$ describing the process `state$~\rightarrow~$decay products$~\rightarrow~$state'
leading to
\begin{equation}
d(E)=\frac{\operatorname{Im}\Pi(E)}{\pi}\left[  (E-M+\operatorname{Re}%
\Pi(E))^{2}+\operatorname{Im}\Pi(E)^{2}\right]  ^{-1}\text{ ,}%
\end{equation}
where $\Gamma(E)=2\operatorname{Im}\Pi(E)$ is the energy-dependent width.
Indeed, $d(E)=-\frac{1}{\pi}\operatorname{Im}[G(E)],$ with $G(E)=(E-M+\Pi(E)+i\varepsilon)^{-1}$
being the propagator of the unstable state. The normalization $\int_{E_{th}%
}^{+\infty}\mathrm{dE}d(E)=1$ is valid in general \cite{lupofermions}.

For the actual calculation of the loop function $\Pi(E)$, one needs a
microscopic model that couples the unstable states to its decay products (it
could be in the form of Lee-Friedrichs Hamiltonian, e.g. \cite{zhou,duecan,mdl}). Often, what is actually known (or assumed) is the function $\Gamma(E)$,
thus $\Pi(E)$ (for complex $E)$ can be reconstructed by the dispersion relation:%
\begin{equation}
\Pi(E)=-\frac{1}{2\pi}\int_{E_{th}}^{\Lambda}\frac{\Gamma(E^{\prime}%
)}{E^{\prime}-E+i\varepsilon}dE^{\prime}+C \text{ ,}
\end{equation}
where $\Lambda$ is a high-energy cutoff (to be sent to $\infty)$ and $C$ is a
real subtraction constant that guarantees that the mass $M$ remains unchanged
(that is, $\operatorname{Re}\Pi(M)=0;$ note, for $E$ real, $\operatorname{Re}%
\Pi(E)$ reduces to the principal part of the integral above).

First, we apply this procedure to the choice $\Gamma(E)=\Gamma\theta(E-E_{th}),$
which is the simplest extension of BW upon introducing a threshold. The
resulting spectral function reads%

\begin{equation}
d(E)=\frac{\Gamma}{2\pi}\left[  \left(  E-M+\frac{\Gamma}{2\pi}\ln\left(
\frac{E-E_{th}}{M-E_{th}}\right)  \right)  ^{2}+\frac{\Gamma^{2}}{4}\right]
^{-1}\text{ }\theta(E-E_{th})\text{ ,}%
\end{equation}
which is correctly normalized, but contains a logarithm in the mass part of the
distribution as well as an unphysical behavior at threshold due to the abrupt
jump of the decay width at that value. Neglecting the $\log$-term would spoil
the normalization and bring back to Eq. \ref{rescal}. In the limit $E_{th}\rightarrow-\infty$ the BW-limit is, as
expected, recovered.

As discussed in \cite{Giacosa:2021mbz}, the simple choice
\begin{equation}
\Gamma(E)=\gamma\sqrt{E-E_{th}}\theta(E-E_{th})
\end{equation}
with $\gamma$ being a constant ($\gamma=\Gamma/\sqrt{M-E_{th}}$), leads to the
nonrelativistic Sill distribution%
\begin{equation}
d^{\text{nrSill}}(E)=\frac{\gamma\sqrt{E-E_{th}}}{2\pi}\left[  (E-M)^{2}%
+\frac{1}{4}\left(  \gamma\sqrt{E-E_{th}}\right)  ^{2}\right]  ^{-1}\text{
}\theta(E-E_{th})\text{ ,}%
\end{equation}
which is correctly normalized to one for any value of the involved parameters
(as long as $M>E_{th}$). Since $\operatorname{Re}\Pi(E)$ vanishes above $E_{th}$, the Sill
does not contain any modification to the `mass part' of the spectral function. Moreover, it describes the left threshold without abrupt jumps. The on-shell
width is obtained as $\Gamma\equiv$ $\Gamma(M)=\gamma\sqrt{M-E_{th}}.$

In order to discuss the relativistic case, we need to perform the replacements
$E\rightarrow s=E^{2}$ (as well as $M\rightarrow M^{2}$ and $E_{th}\rightarrow
s_{th}\geq0$ for the lowest energy threshold), and $\Pi(E)\rightarrow\Pi(s)$
with $\operatorname{Im}\Pi(s)=\sqrt{s}\Gamma(s).$ Thus, in analogy to the QM
case, an unstable resonance is described by
\begin{equation}
d(s)=\frac{\operatorname{Im}\Pi(s)}{\pi}\left[  (s-M^{2}+\operatorname{Re}%
\Pi(s))^{2}+\operatorname{Im}\Pi(s)^{2}\right]  ^{-1}\text{ ,}%
\end{equation}
which is correctly normalized, $\int_{s_{th}}^{+\infty}\mathrm{dE}d(s)=1$. Moreover, $d(s)=-\frac{1}{\pi}\operatorname{Im}[G(s)]$
with the relativistic propagator
$G(s)=(s-M^{2}+\Pi(s)+i\varepsilon)^{-1}$.
The loop function $\Pi(s)$
reads
\begin{equation}
\Pi(s)=-\frac{1}{\pi}\int_{s_{th}}^{\Lambda^{2}}\frac{\sqrt{s^{\prime}}%
\Gamma(s^{\prime})}{s^{\prime}-s+i\varepsilon}dE^{\prime}+C
\end{equation}
where $C$ is chosen such that $\operatorname{Re}\Pi(M^{2})=0,$ thus the
nominal mass is left unchanged.

The relativistic Sill distribution corresponds to the choice $\Gamma(s)=\tilde{\Gamma}\sqrt{(s-s_{th})/s}$, leading to
\begin{equation}
d^{\text{Sill}%
}(s)=\frac{\tilde{\Gamma}\sqrt{s-s_{th}}}{\pi}\left[  (s-M^{2})^{2}%
+\tilde{\Gamma}^{2}\left(  s-s_{th}\right)  \right]  ^{-1}\theta
(s-s_{th})\text{ }.
\label{eq:Sill}
\end{equation}
It is normalized to unity for any choice of $M$ and $\tilde{\Gamma}$, the left-threshold is taken into account,
$\operatorname{Re}\Pi(s)=0$ above $s_{th}$, thus making it quite simple to use. Moreover, the
width function $\Gamma(s)$ reduces to a constant for large values of $s,$ thus
heuristically one may consider the Sill distribution as the proper extension
of the BW\ function to the relativistic case. This is also evident in the
limit $s_{th}=0,$ for which $\Gamma(s)=\tilde{\Gamma}$ is a constant.
\begin{figure}
    \centering
    \includegraphics[width=0.8\textwidth]{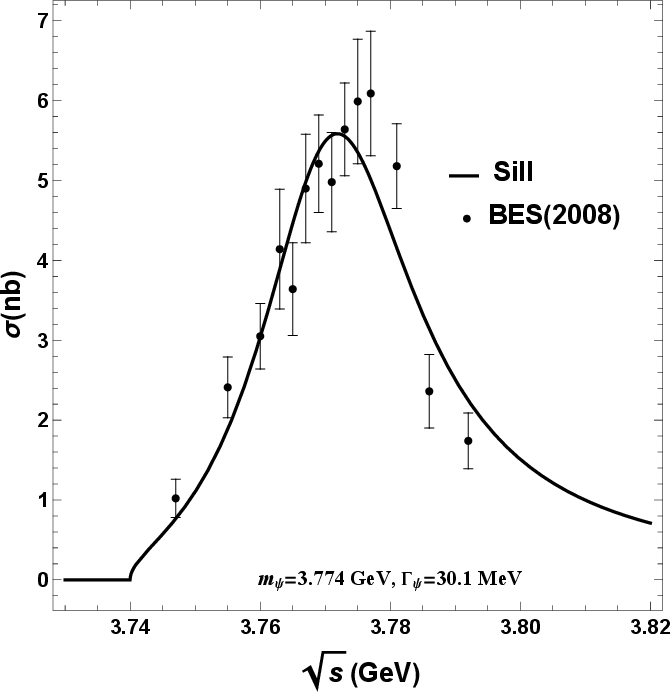}
    \caption{The $e^+e^-\to D\bar{D}$ data reported by BES (Table 2 of Ref.~\cite{BES:2008rkr}) fitted with the Sill distribution ($\chi^2$ per d.o.f is $1.59$). The mass and width of $\psi(3770)$ obtained from the fit are shown at the bottom of the figure.}
    \label{fig:3770fit}
\end{figure}


Various comments are important at this point:\par

1) For a two-body decay into particles with masses $m_1$ and $m_2$, one sets $s_{th}=(m_{1}+m_{2})^{2}.$ In general, the Sill is
different from the Flatt\'{e} distribution \cite{flatte} (see also e.g. \cite{baru,rui}) for which $\operatorname{Im}%
\Pi(s)\sim k(s)$ with $k(s)$ being the momentum. 

2) Extension to the multichannel case and to decay chains is straightforward \cite{Giacosa:2021mbz}. The Sill can be applied to both mesons and baryons.

3) One may obtain the relativistic BW (rBW) distribution by setting
$\operatorname{Im}\Pi(s)=M\Gamma\theta(s)$ (a left threshold must exist
in\ QFT). Interestingly, this corresponds to a non-constant width of the type
$\Gamma(s)=\Gamma M/\sqrt{s}.$ The resulting distribution 
\begin{equation}
d(s)=\frac{\Gamma M}{\pi}\left[  \left(  s-M^{2}+\frac{\Gamma M}{\pi}%
\ln\left(  s/M^{2}\right)  \right)  ^{2}+\left(  \Gamma M\right)  ^{2}\right]
^{-1}\theta(s) 
\end{equation}
reduces to $d^{\text{rBW}}(s)=N\frac{\Gamma M}{\pi}\left[  \left(
s-M^{2}\right)  ^{2}+\left(  \Gamma M\right)  ^{2}\right]  ^{-1}\theta(s)$
when the log-term is neglected (see e.g. \cite{pythia} for applications).

4) Although all distributions are peaked at the mass $M$, the high-$E$ scaling are
quite different. BW scales as $E^{-2}$ , the nrSill goes as $E^{-3/2}$ , the
rBW as $E^{-3}$ and the Sill as $E^{-2},$ as BW.
(Note, due to the variable change $s=E^2$, one has $d^{\text{Sill}}(E) =2Ed^{\text{Sill}}(s=E^2)$).

5) In the complex plane, the Sill loop $\Pi(s)=i\tilde{\Gamma}\sqrt{s-s_{th}}$
in the first Riemann sheet (RS) contains a $(s_{th},\infty)$-cut. In the II
RS, $\Pi_{II}(s)=\Pi(s)+2i\tilde{\Gamma}(\sqrt{s-s_{th}})_{II}=-i\tilde{\Gamma}%
\sqrt{s-s_{th}}$ has a similar feature. While BW contains only one sheet, the Sill
has a richer and more realistic (although not complete, see below) analytic structure.

6) An actual scalar QFT of the type $\mathcal{L}=gS\varphi^{2}$ leads to the
decay width $\Gamma(s)=\frac{g^{2}\sqrt{s-s_{th}}}{8\pi s}$ \cite{lupo}, where $s_{th}=4m^2$, with $m$ being the $\varphi$ mass. This expression has different analytical properties with respect to the
Sill in Eq.~\ref{eq:Sill}.  The self-energy takes the form:
\begin{equation}
\Pi(s)=\frac{g^{2}\sqrt{s-s_{th}}}{8\pi\sqrt{s}}\ln\left(  \frac
{\sqrt{s-s_{th}}-\sqrt{s}}{\sqrt{s-s_{th}}+\sqrt{s}}\right)  +C\text{ .}%
\end{equation}
In the I Riemann sheet it has a cut on the real axis that extends from $s_{th}$ to
$\infty$ (note, $s<0$ is, thanks to the log-function, regular). In the II RS,
$\Pi_{II}(s)=\Pi(s)+2i\frac{g^{2} \left( \sqrt{s(s-s_{th})}\right)_{II}}{8\pi s}$, which
contains also a branch cut on the negative $s$-axis, a feature not
contemplated in the Sill. In this sense, the Sill simplifies certain analytic
properties, but still retains the most important ones.

In Ref. \cite{Giacosa:2021mbz} various examples of resonances described via the Sill were presented. Here, as a novel one, the line shape of the charmonium
resonance $\psi(3770),$ with one dominating $\ \bar{D}D$ decay channel, is plotted in Fig.~\ref{fig:3770fit},. The mass and width of $\psi(3770)$ are obtained as $m=3.774\text{ GeV}$ and $\Gamma = 30.1\text{ MeV}$ respectively, in good agreement with the PDG values. To arrive at these values, we fit the Sill distribution the data reported by BES collaboration (Table 2 of Ref. \cite{BES:2008rkr}) along with a normalization constant that takes care of the phase space factors. Quite remarkably, the Sill fits as well as the complete loop treatment of this resonance discussed in \cite{Coito:2017ppc} (to which we refer for connecting the cross section to the spectral function) \par

Meanwhile, the Sill has been already used
in the review paper for the description of the $\Delta$ baryon \cite{Winney:2022tky} as well
as by the Alice collaboration \cite{ALICE:2023wjz} for the description of the baryon state
$\Xi(1620)$  (see also \cite{Sarti:2023wlg}).
In the future, one may search for Sill extensions that are better
suited to describe higher angular momentum waves \cite{peters} and to three-body decay rates (see, Ref.~\cite{Jafarzade:2023acz} for a phenomenological exposition).

\bigskip

\textbf{Acknowledgments} We thank A. Okopi\'nska, A. Pilloni, and M.~F.~M. Lutz for useful discussions. Financial support  from the Polish National Science Centre (NCN) via the OPUS project
2019/ 33/B/ST2/00613 is acknowledged.

\end{document}